# Quantum Imaging of Single-Atom Spin-Splitting in a Monolayer Semiconductor

Caleb Z. Zerger[1,2], Alex W. Contryman[1,2], Changmin Lee[3,4], Shreyas Patankar[3,4], Joseph Orenstein[3,4], Tyler J. Layden[5], Marc A. Kastner[1,5], David Goldhaber-Gordon[1,5], Xiaolin Zheng[6], Hong Li[6,7]*, Hari C. Manoharan[1,5]*

[1]Stanford Institute for Materials and Energy Sciences, SLAC National Accelerator Laboratory, Menlo Park, CA 94025, USA

[2]Department of Applied Physics, Stanford University, Stanford, CA 94305, USA

[3]Materials Science Division, Lawrence Berkeley National Laboratory, Berkeley, CA 94720, USA

[4]Department of Physics, University of California, Berkeley, CA 94720, USA

[5]Department of Physics, Stanford University, Stanford, CA 94305, USA

[6]Department of Mechanical Engineering, Stanford University, Stanford, CA 94305, USA

[7]School of Mechanical & Aerospace Engineering, Nanyang Technological University, Singapore 639798

*To whom correspondence should be addressed (ehongli@ntu.edu.sg and manoharan@stanford.edu)

**Abstract:**

Theoretical work (13, 14) has suggested that monolayer $MoS_2$ doped with Mn should behave as a two-dimensional dilute magnetic semiconductor, which would open up possibilities for spintronic applications, device physics, and novel ground states (4, 14). The magnetic properties on Mn dopants in $MoS_2$ are dependent on the mid-gap impurity states of said dopants as well as the sites of dopant incorporation and dopant concentration. In this work we use STM/STS to characterize multiple impurity types associated with Mn dopants in $MoS_2$, and use ring features that appear in spectral maps due to tip-induced band bending to investigate the nature of the mid-gap impurity states. The doublet nature of the rings and comparison to DFT calculations show that the Mn states exhibit strong spin splitting which can be quantified. We used scanned MOKE experiments to extend these magnetization measurements from atomic scale to mm scales, and detect the spin susceptibility signal which increases with Mn concentration. These experiments show that single Mn atoms in $MoS_2$ function as active unscreened magnetic moments in the TMD monolayer, and can be harnessed for spin physics applications and science.

**Introduction:** Since the discovery of ferromagnetism in Mn doped GaAs (1) researchers have been interested in the potential applications of dilute magnetic semiconductors (DMS). Using these materials, researchers hope to create so-called "spintronic" devices in which the coupled semiconducting and magnetic properties of DMSs are used to generate spin-polarized current which can be used in applications such as Magnetic-RAM (2-4). However, all known DMSs have magnetic ordering temperatures well below room temperature, making it difficult to use them for relevant applications.

Monolayer $MoS_2$ is a semiconducting transition metal dichalcogenide (TMD) which has been researched extensively due to its novel electronic and optical properties (5-11). Some recent theoretical work suggests that MoS2 can become a DMS when doped with magnetically active transition metal atoms such as manganese. The DMS state arises from the coupling of manganese atoms substituted for molybdenum atoms through indirect exchange at sufficiently high concentrations, in the regime of 5-10% (13, 15-17). Some predicted Curie temperatures are in the range of 1000K, making these materials attractive for spintronics applications (14). Scanning tunneling microscopy (STM) and spectroscopy (STS) are powerful tools for studying doped semiconductors, as they are well equipped to probe and manipulate semiconductor properties on the length scale of an individual dopant (22-23).

A common feature of STM/STS measurements on doped semiconductors is the presence of ring features which appear around impurities in spectral maps (21). These rings occur when the tip is at a distance from the impurity such that tip-induced band bending (TIBB) aligns an impurity state with the Fermi energy (18,20). This causes the impurity state to depopulate, resulting in a change of tunnel current at some voltage-dependent radius from the dopant. This shows up on an energy map as a ring, which varies in size as the energy is varied. In this article, we present observation of ring features in the spectral maps taken of Mn doped $MoS_2$ using STM/STS. In particular, we observe prominent ring doublets around specific Mn sites, with rings appearing and dispersing at different energies. We find that the number of rings and energies at which they appear is associated with the geometry of the impurity as seen in the topography. Taken in conjunction with previous density functional theory (DFT) calculations on Mn-$MoS_2$, our STM/STS observations suggest a large energy splitting between impurity states of Mn dopants, which we associate with known magnetic states of Mn dopants incorporated into the monolayer $MoS_2$ lattice. Scanning Kerr measurements over large areas confirm that the Mn atoms are magnetically active, and their spin susceptibility increases with local Mn concentration.

**Results and Discussion**: Our STM measurements were performed at liquid helium temperature (4.2K) on a manganese-doped monolayer $MoS_2$ sample grown by chemical vapor deposition on an $SiO_2$ substrate with a gold top contact (Figure 1a). A typical STM topograph ($V_s$ = -2.7 V, $I_s$ = 167 pA, shown in Figure 1c) shows several impurity types which we will explore in detail: we will refer to these impurities as dark spots (labeled in yellow in Figure 1c), Mo substitution (labeled in green), disulfur substitution (labelled in red), and 2 lobed (labelled in blue). We later present more details to justify these assignments of Mo substitution and disulfur substitution. The concentration of all impurities over this region is about 6 x $10^{12}$ $cm^{-2}$. In a spectral map like that shown in Figure 1d we can associate each impurity type with dark rings that appear when the TIBB depopulates an impurity state, leading to a sharp drop in conductance and thus a dark ring in the spectral map (30). For each impurity type we see rings which appear at different energies and grow with energy at different rates (see Supplementary Movies showing animated constant-energy slices through the entire conductance map dataset).

In Figure 2 we show close-up topography and typical STS on our samples. Far from impurities (Figure 2 a-c) the sample looks like monolayer $MoS_2$, with STS away from impurities exhibiting a bandgap of 2.3 eV, with the conduction band minimum (CBM) 0.25 eV above the Fermi energy, suggesting an n-doped sample consistent with past STS studies of monolayer $Mo_2$ (24). The dark spot impurity (Figure 2d-f) appears as a single ring in the spectral maps, and the STS shows a shift in the valence band maximum (VBM) up to -1.5 V. This is consistent with a trapped charge in a substrate defect seen in other STM work (26).

The disulfur substitution (Figure 2g-i) shows multiple rings appearing at different energies. In the STS we can see two sharp conductance minima corresponding to two of these dark rings that arise from the depopulation of impurity states by TIBB. As we will explore later, this is consistent with DFT calculations for in-gap states of a Mn dopant substituted for a disulfur vacancy in the $MoS_2$. An Mn disulfur substitution is predicted to be magnetic, with the in-gap impurity states demonstrating strong spin character. We see similar features for the Mn impurity substituted on an Mo site. We first note the threefold symmetry of the topography of these impurities (Figure 2j), and that this threefold symmetry is 180° rotated from that of the disulfur impurity (Figure 2g), which we expect for dopants on different sublattices. The spectral maps/STS (Figure 2k,l) show the TIBB rings, again a doublet closely spaced in location and energies. We see from a typical STS on this impurity two dips associated with two rings, suggesting we are accessing two in-gap impurity states. We show that this picture is consistent with DFT work done by Ramasubramaniam et. al (14) who calculates two spin-polarized in-gap states below the Fermi level in Mn doped $MoS_2$, one with a strong spin up character and the other spin down. These states are of particular interest, as these calculations also suggest that at sufficiently high concentration Mn doped $MoS_2$ exhibits long range ferromagnetism, as it forms a 2D DMS.

In Figure 3 we look more closely at the development of the TIBB rings. Supplementary Figure 2b-d shows a full video of ring growth for each impurity type, but we here highlight energies demonstrating important features. The dark spot impurity shows a single ring which grows only slightly before disappearing. The Mn impurity shows two rings; the first disappears near $V_s$ = -2.0 V at which point another ring appears, growing in size before disappearing. For the disulfur substitution we can see at $V_s$ = -2.7 to -2.1 V two concentric rings, suggesting two impurity states close in energy, consistent with the picture calculated by Cong et. al as described above (25). At $V_s$ = -1.1 V we can see a third ring form.

All of these features can be conveniently summarized in the plots shown in Figure 4, where we plot color coded *dI/dV* spectra as a function of energy and distance from the dopants. We can see that for the dark spot impurity (Figure 4a), the dark ring has a radius of about 1 nm over the entire energy range that it appears. Because this ring appears when the band bending is sufficient to align an impurity state with the Fermi level, the dark rings as shown in the color coded spectra correspond to a line of equal band bending, which we mark with a dashed red line. We note that these rings grow as $V_s$ is increased, suggesting the flatband condition corresponds to $V_s$ much less than -3 V. For the Mo substitution case (Figure 4b) we mark the two rings, and note that the *dI/dV* is highest directly over the impurity, suggesting a charging of the impurity as it becomes depopulated due to TIBB.

We construct a simple model for this similar to that used in past work on doped 3D semiconductors (20) but using a 2D Poisson solver with a single impurity which charges when the surface potential is raised above a threshold level, $E_1$, necessary to raise the impurity state to the Fermi level $E_F$, and repeat this process for a second charging state at energy $E_2$. We incorporate this into a canonical Tersoff-Hamann model to calculate the expected differential conductance. We show in Figure 4d-e that for typical STM conditions the ring we observe at $V_s$ = -2.14 V for the Mo substitution dopant is best fit in our model with an impurity state $E_1$ = -0.95 eV. A second ring of radius *r* = 0.5 nm for $V_s$ = -1.0 V (similar to what we observe in our system) is calculated for a charging energy of $E_2$ = -1.55 eV. We note that these energies are further from the Fermi level than predicted by DFT (14); however, the Fermi level is shifted significantly further from the VBM in our samples than in the DFT calculations. While the energy of the impurity levels is shifted relative to DFT the spin-splitting $E_1 - E_2$ = 600mV is consistent with the DFT. Another parameter is that this calculation is dependent on the assumption of the flat band condition, which lies deep in the valence band and from the "volcano" plots in Figure 4 was deduced to be $V_{FB}$ = -4V. The last color coded spectra (Figure 4c) shows the three dark rings associated with the states of a disulfur substitution, where we mark each of the three rings shown in Figure 3. The two rings which appear at similar energies are consistent with the 3*d xy/yz* spin split state of an Mn disulfur substitution with a spin splitting $\Delta E$ = 30 mV as calculated by Cong et. al. (25), as shown by calculations using the Poisson solver model described above.

We note that while our model suggests our data is consistent with the impurity levels calculated in the DFT work cited above (14, 25) DFT calculations vary by hundreds of millivolts based on the functional chosen, the impurity concentration enforced by choice of grid, etc., and tend to underestimate the band gap. We also note that the low CBM suggests our sample is n-doped, which could also be relevant to impurity levels. For example, Xianqing Lin, et al. suggest that for higher chemical potential, a charge state of *q* = -1 or *q* = -2 could be favored for Mn substitutional dopants in $MoS_2$, and that these charge states are associated with higher magnetic moments for the dopant (27), but are nonetheless still spin polarized. While further work is necessary to conclude which specific charge states are being probed by measuring the TIBB rings, the basic picture of magnetically active Mn dopants in $MoS_2$ remains the same.

In the Mn-$MoS_2$ samples studied by STM above, the concentration of dopants was <0.5%, lower than the concentration at which long range ferromagnetic ordering is expected to be present. We provide further evidence for magnetically active free spins in Mn-doped $MoS_2$ samples at variable concentrations via AC magneto-optic Kerr effect (MOKE) measurements as shown in Figure 5. Here the Kerr rotation was measured while scanning over an Mn-$MoS_2$ sample from a region of low concentration of Mn to regions of higher concentration (the gradient was induced by the growth and increases along the +Y measurement direction). In Figure 5a this measurement was acquired at 5K, and a clear increase in spin susceptibility through the ac Kerr signal is shown with increasing Mn concentration, while at 100K (Figure 5b) no such trend is observed. The subtraction of these two data sets shows clear evidence of a spin signal tied to the Mn doping level.

**Conclusions**: In summary, we have observed multiple types of impurities in Mn doped $MoS_2$. These impurities show evidence of multiple impurity states above the valence band edge and below $E_F$ which we probe by measuring the rings formed due to TIBB. Taken in conjunction with DFT calculations, our findings exhibit signatures that Mn dopants are magnetic in $MoS_2$. This conclusion is supported by MOKE data over larger length scales showing that at higher concentrations of Mn these samples exhibit a finite and increasing spin susceptibility. The presence of spin-split Mn atoms incorporated into $MoS_2$ is therefore experimentally shown to be a promising candidate system for a 2D DMS.

**Acknowledgements:** This work is supported by the U.S. Department of Energy, Office of Science, Basic Energy Sciences, Materials Sciences and Engineering Division, under Contract DE-AC02-76SF00515.

**Methods:**

**Sample growth of Mn-doped monolayer MoS$_2$**

Molybdenum trioxide (MoO$_3$), Manganese sulfide (MnS), and sulfur (S) powder were used as sources. 5 mg MoO$_3$ powder was loaded in a ceramic boat. A piece of Si wafer capped with 270 nm SiO$_2$ layer was suspended on the ceramic boat with the polished side facing up. Along the edge of the Si wafer piece, MnS (2 mg) particles were spread evenly as much as possible. Then the ceramic boat was located at the center of a quartz tube furnace. Another ceramic boat containing sulfur powder (1 g) was located at an upstream position where the temperature would reach 200 ºC during the growth. After sealing, the tube was pump down to 0.01 mbar, and then 100 sccm argon (Ar) gas was flowed in till atmosphere pressure. This procedure was repeated for 3 times to remove the oxygen molecules in the tube as much as possible. Then heated to 850 ºC within 15min, and kept at 850 ºC for 20 min. The system was then cooled down to room temperature naturally. (See Supplementary Figure 1 for growth schematic).

The density of Mn-doped MoS2 flakes decreases gradually from edge to the centerline of the wafer as the MoO3 vapor concentration decreases from the edge towards the center of the wafer that has similar size as the alumina boat (10-mm-wide and 50-mm-long). Raman characterization across the wafer surface shows majority of the film consists of monolayer MoS$_2$ in the center region of the wafer. The evaporated MnS vapor transports along with MoO$_3$ vapor to introduce Mn dopants in the grown MoS$_2$.

**Sample Preparation**

To minimize the contamination from process, a shadow mask was employed to form metal contact to Mn-doped MoS$_2$ sample. The shadow mask was a TEM grid purchased from Ted Pella Inc. A titanium (5 nm)/gold (50 nm) bilayer was deposited using an E-beam evaporator onto the sample covered by the shadow mask. The metallic contact pads formed also serve as a mark to locate the Mn-doped MoS$_2$ flakes that are connected to the metallic contact pads.

**STM/STS**

Measurements were performed under UHV at 4 K in a UHV STM using a Pt-Ir tip, which was first checked for spectral flatness and cleanliness on Au(111). *dI/dV* spectra were acquired using a standard lock-in technique at 587 Hz and with typical modulation voltage of 14 meV, and spectral maps were acquired in runs of ~12 hours with typical energy steps of 50 meV and spatial resolution of 1.0 Å. Topographies were acquired in constant current mode.

**Modeling TIBB rings**

TIBB rings were modeled following the method of Marczinowski et. al. (20). We first calculated a surface potential using a 2D Poisson solver modeled after the method used in SEMITIP software (28) but adapted to a 2D semiconductor with a single chargeable impurity. We use this to find impurity levels $E_1$ and $E_2$ that best reproduce our data. Values for relevant monolayer MoS$_2$ properties such as dielectric constant were taken from Hill et. al. (29) and a tip distance of 5 Å was assumed based on measurement parameters. The band bending calculated from this Poisson solver and potential from the charged dopant inside the charging radius was then integrated into a standard Tersoff-Hamann model in order to calculate the expected *dI/dV*.

**Magneto-optic Kerr effect (MOKE) measurements**

In this AC magneto-optic Kerr effect (MOKE) microscopy setup, a linearly polarized 633 nm HeNe laser beam was focused at normal incidence onto a 1 µm spot on the sample surface with an objective lens (Olympus LMPFLN 50x, NA = 0.5). The reflected beam was then collected by a 50:50 non-polarizing beam splitter, which directs the beam to a quarter wave plate and a Wollaston prism for balanced photodetection of Kerr ellipticity. The sample position was raster-scanned with *xy* piezoelectric scanners (Attocube ANPx101) and the focusing was fine-adjusted with a *z* piezoelectric scanner (Attocube ANPz102). An out-of-plane AC magnetic field was applied through a coil (Woodruff Scientific, 156 turns, inner diameter: 5mm, height: 1.5 mm) that surrounds the sample.

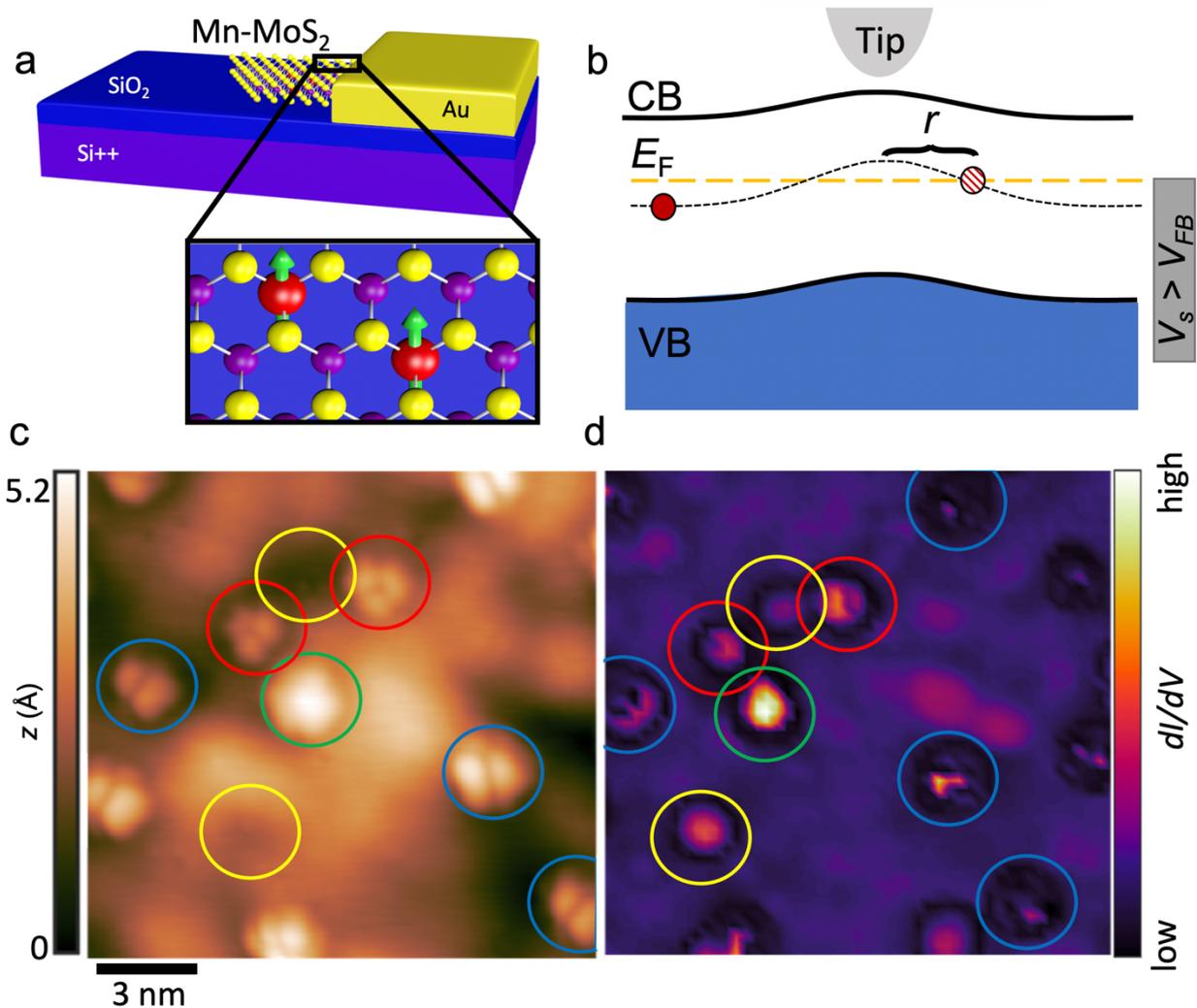

**Figure 1:** a) Sample schematic shows Mn-doped monolayer $MoS_2$ on silicon dioxide with a gold top contact. Mn substituted on Mo sites is predicted to give rise to magnetism. b) Illustration of tip-induced band bending rings. When the sample bias $V_s$ is above the flatband condition the bands will bend up near the tip. When this is at a distance $r$ sufficient to bring an in-gap impurity state level (hatched circle) with the Fermi level $E_F$, there is a resonance effect in the current, and when this channel is closed off due to increased $V_s$, there is a dip in conductance as the state is depopulated. This manifests as a dark ring in the spectral map. For contrast, another dopant state (filled red circle) that is still occupied is shown at a distance greater than $r$. c) Constant-current topography ($V_s$ = -2.3V, $I_s$ = 167 pA) taken over a 15nm window. We observe four different types of Mn impurities: i) Circled in red: Single Mo substitutional Mn dopant has a three lobed structure which aligns with the underlying hexagonal $MoS_2$ lattice. ii) Circled in green: This feature is a disulfur substitutional Mn dopant which has the opposite threefold rotational symmetry as the Mn substitution. iii) Circled in blue: Two-lobed feature, possibly due to adjacent dopants. iv) Circled in yellow: The dark spot impurities are likely a substrate defect leading to trapped charge. d) Associated $dI/dV$ map for the topography in a). This map represents a constant-energy differential conductance map where we observe TIBB rings around the various types of impurities.

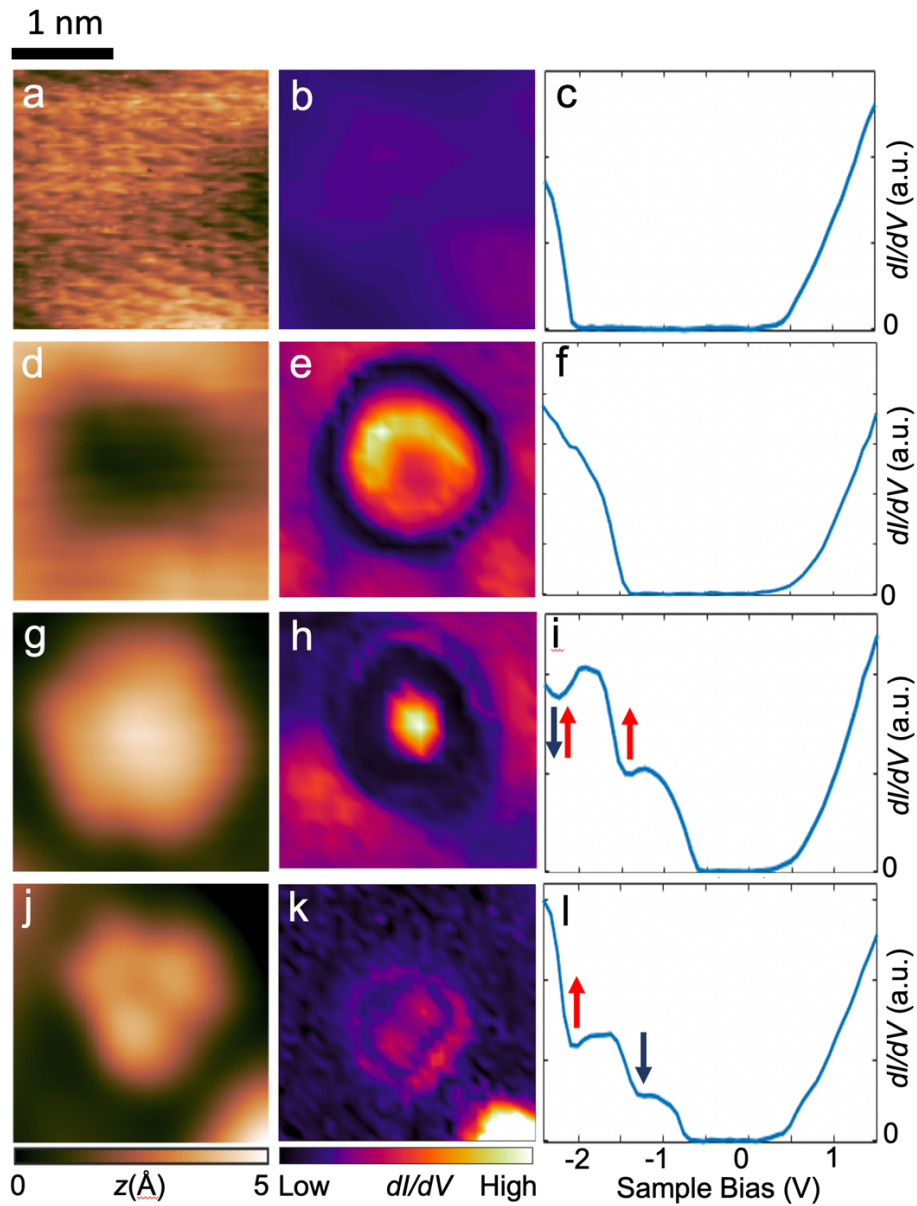

**Figure 2:** a) Constant current topography of bare MoS$_2$ (V$_s$ = -2.7V, I$_s$ = 167 pA). Scan size is 2.5nm. b) An energy slice of the full *dI/dV* map at -2.5V over same bare MoS$_2$ and c) typical STS over bare MoS$_2$. d) Topography of dark spot impurity, showing in e) one ring in the *dI/dV* map energy -2.5V. f) STS taken near center of impurity shows an upward shift in valence band maximum. g) Topography of Mn as a disulfur substitutional dopant, showing threefold coordination. h) *dI/dV* map at energy -2.14V shows two simultaneous rings. i) STS spectra showing two mid-gap bumps and dips due to charging of impurity states with proposed associated spin character. j) Topography of Mn as Mo substitutional dopant, again showing threefold coordination. k) *dI/dV* map at -1.4 V showing a ring doublet. l) STS spectra where mid gap bumps and dips due to charging of the impurity states is associated with two strongly spin-split impurity states.

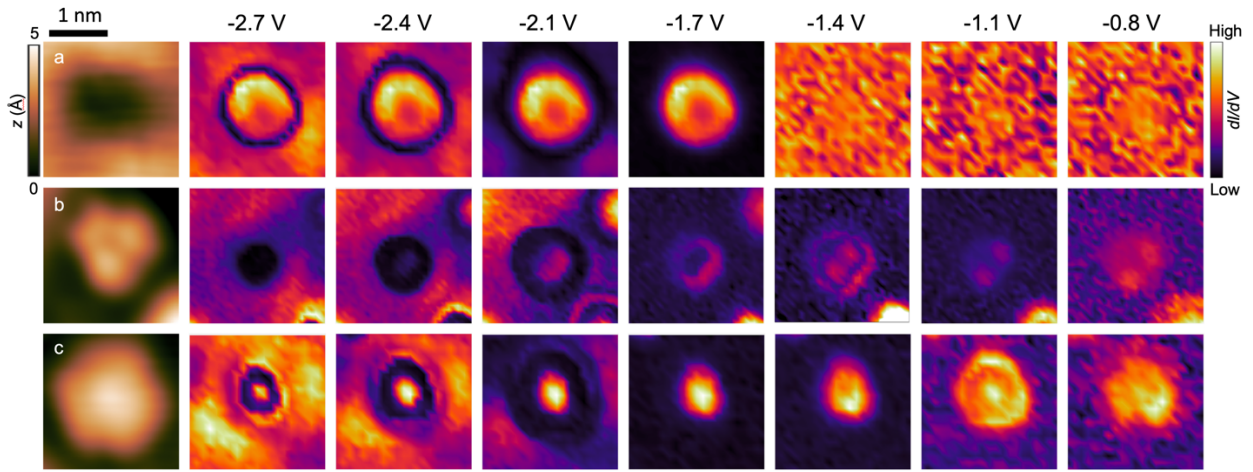

**Figure 3:** a) Topography and *dI/dV* maps for a dark spot impurity, showing a single ring. b) Shows the topography and *dI/dV* maps for an Mo substitutional Mn dopant. The *dI/dV* maps show two rings, that grow in size with increasing bias. The second ring first can first be seen at -1.7 V. c) Shows the topography and *dI/dV* maps for an $S_2$ substitutional dopant, showing two rings that grow and disappear around -1.7V, and a third that appears around -1.4V.

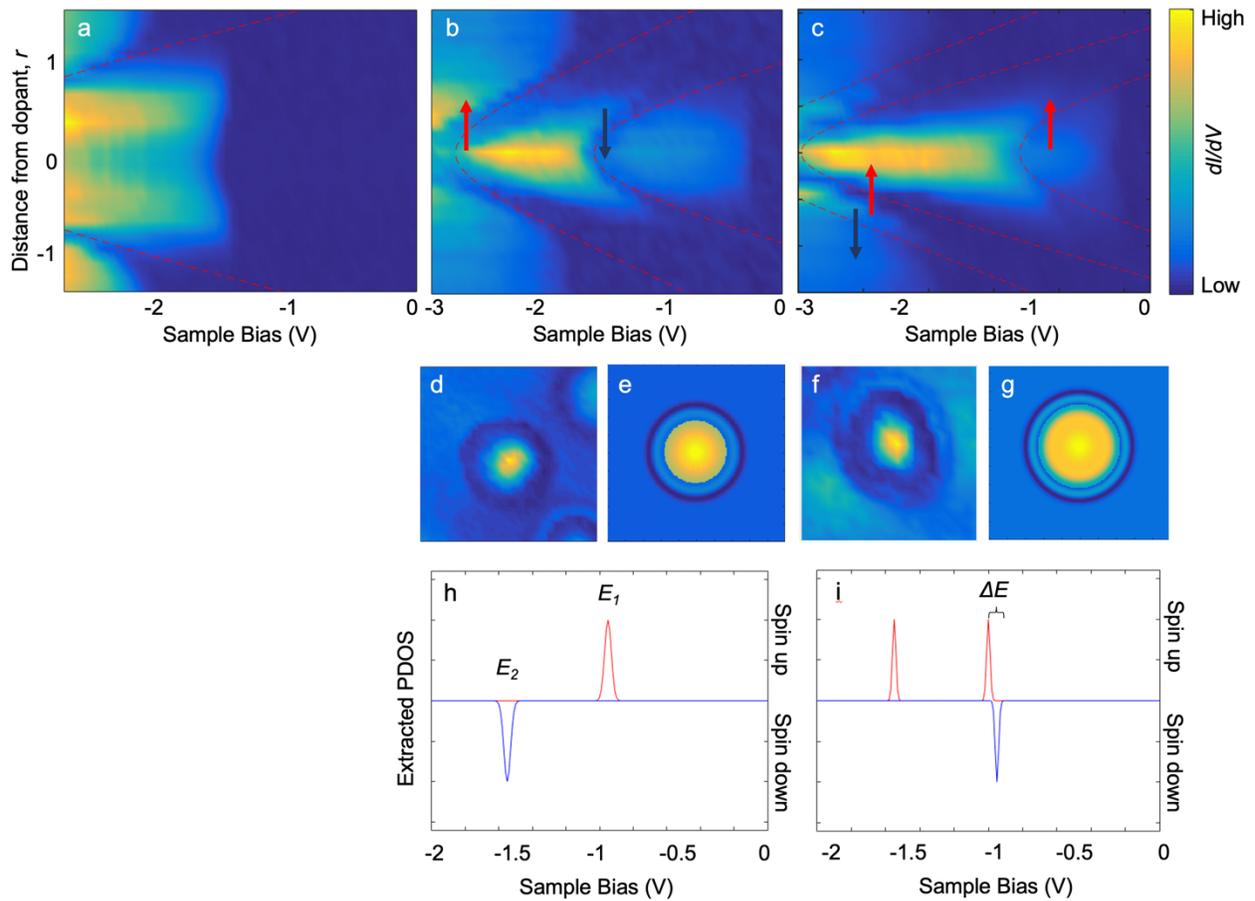

**Figure 4:** Color-coded *dI/dV* spectra over a) the dark spot impurity b) the Mo substitutional dopant and c) the $S_2$ substitutional dopant. Dark rings correspond to lines of equal TIBB, marked with red dashed lines. Arrows denote deduced spin character of charged impurity states. d) *dI/dV* map over Mo substitution at -2.14V, next to e) a simulated *dI/dV* map at at -2.14V and typical STM conditions, with the model simulated for a midgap state at energy $E_1$ =-0.95 eV. One ring is visible here. f) *dI/dV* map of a disulfur substitution at -2.14V, and g) simulated *dI/dV* map at -2.14V and a spin splitting of ΔE = 30 mV. h) shows a simplified PDOS extracted for Mn as an Mo substitutional dopant. The difference $E_1 - E_2$ = 600 meV is consistent with DFT calculations (14) i) Shows the same for the $S_2$ substitutional dopant, where ΔE = 30 mV for a single spin-split impurity state as predicted by DFT (25).

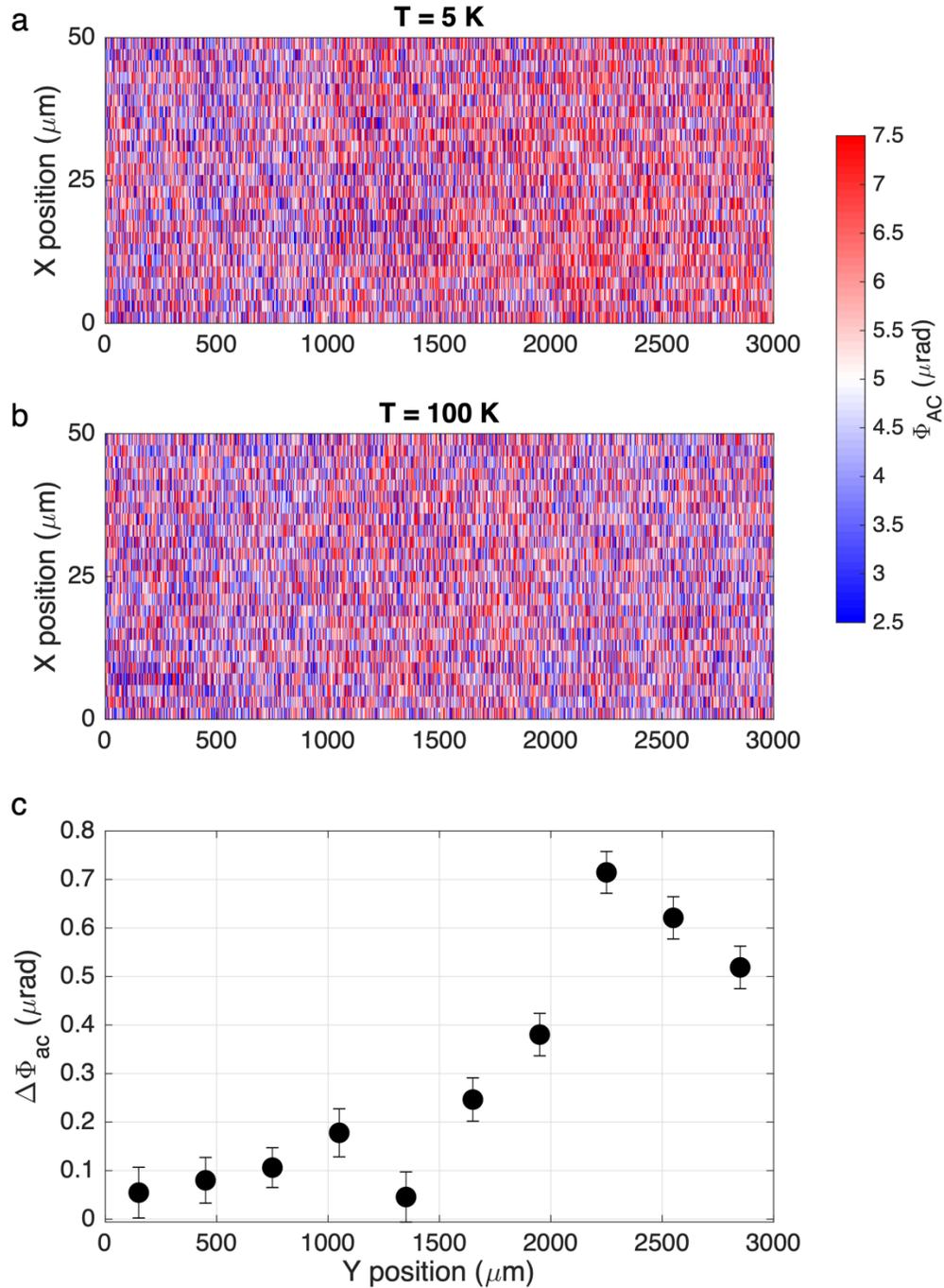

**Figure 5:** Scanning MOKE microscopy measurements. a, Color scale map of the Kerr effect amplitude in a 50 x 3000 μm region shows a gradual increase in the signal intensity toward denser Mn concentration, in agreement with the growth-induced gradient Mn-doping of the sample which increases along the +Y direction. The measurements were taken at $T$ = 5 K with a step size of 2 μm. b, The same measurement taken at $T$ = 100 K shows little change in signal amplitude across the sample. c, The average value of the change in Kerr effect between 5 and 100 K is plotted as a function of location along the strip mapped out by measurements in a and b. Each datapoint is averaged over contiguous regions of 50 x 300 μm and the error bars correspond to the standard deviation of the mean from these measurements.